\documentclass[useAMS,usenatbib]{mn2e}

\usepackage[dvips]{graphicx}
\usepackage[]{txfonts}
\def\simless{\mathbin{\lower 3pt\hbox
{$\rlap{\raise 5pt\hbox{$\char'074$}}\mathchar"7218$}}}   %< or of order
\def\simmore{\mathbin{\lower 3pt\hbox
{$\rlap{\raise 5pt\hbox{$\char'076$}}\mathchar"7218$}}}   %> or of order
\newcommand{\be}{\begin{equation}}
\newcommand{\ee}{\end{equation}}
\title[UHECRs from magnetic reconnection in jets]{UHECRs from magnetic reconnection in relativistic jets}

\author[Dimitrios Giannios]
{Dimitrios Giannios\thanks{E-mail: giannios@astro.princeton.edu}\\
Department of Astrophysical Sciences, Peyton Hall, Princeton
  University, Princeton, NJ 08544, USA\\}

\begin{document}
\date{Received / Accepted}
\pagerange{\pageref{firstpage}--\pageref{lastpage}} \pubyear{2009}

\maketitle

\label{firstpage}

\begin{abstract}

Ultra-high-energy cosmic rays (UHECRs) may be produced
in active galactic nuclei (AGN) or gamma-ray burst (GRB) jets.    
I argue that magnetic reconnection in jets can accelerate
UHECRs rather independently of physical processes in the 
magnetic dissipation region. First order Fermi acceleration
can efficiently take place in the region where the 
unreconnected (upstream) magnetized fluid converges into
the reconnection layer. I find that protons can reach energies up to 
$E\sim 10^{20}$ eV in GRB and powerful AGN jets
while iron nuclei can reach similar energies in AGN jets of more moderate
luminosity.

\end{abstract} 
  

\section{Introduction} 
\label{intro}

The origin of UHECRs (particles with energy $E\simmore 10^{19}$ eV)
remains a mystery. The fact that magnetic fields cannot confine them
in the Galaxy and the observed spectral cutoff at $\sim 6\times 10^{19}$ eV point to 
extragalactic sources at these energies (Abbasi et al. 2008; Abraham
et al. 2008). For any astrophysical source to accelerate particles to the highest
observed energies of $E_M\simmore 10^{20}$ eV, tight constraints need to
be satisfied. Possibly, the tightest come from the need for particles
with energy $E_M$ to be confined within the size of the source 
(Hillas 1984) and the acceleration to take place within the available (dynamical) time of
the system. Both constraints may be met in very powerful sources.
Viable astrophysical sources for UHECRs are relativistic jets from 
active galactic nuclei (see,.e.g. Halzen \& Hooper 2002) 
and gamma-ray bursts (Milgrom \& Usov 1995; Waxman 1995;Vietri 1995; 
see, however, Ghisellini et al. 2008; Inoue 2008 for alternatives).

The acceleration mechanism for UHECRs is usually postulated to be the first order
Fermi mechanism in (mildly) relativistic shocks in the jets (because of internal or external
interactions; Gallant \& Achterberg 1999; Achterberg et al. 2001). 
For the UHECRs to be confined by shocks both the upstream and
downstream need to be strongly magnetized (i.e. with magnetic energy
density not much less than the total energy density; e.g. Waxman
1995). The magnetic fields cannot be of
small scale (generated by plasma instabilities) since in this case the small angle scattering of the
energetic particles makes the acceleration process too slow to be of
relevance (Kirk \& Reville 2010). Strong, large scale fields are also unlikely to work since
they do not allow for the particles to repeatedly cross the shock 
front (where the acceleration takes place) for most field inclinations 
(e.g. Sironi \& Spitkovsky 2009).

Here I discuss an alternative mechanism for UHECR acceleration in a
relativistic flow. This mechanism applies to magnetic reconnection
regions in Poynting-flux dominated flows (e.g. flows with
Poynting-to-kinetic flux ratio $\sigma\simmore 1$). 
The particle acceleration takes place through the first order
Fermi mechanism because of particles reflected in the magnetized plasma
that is converging in the reconnection region (see Speiser 1965 and
de Gouveia dal Pino \& Lazarian 2005 for a similar
mechanism applied to Earth's magnetotail and the ejections from the
microquasar GRS 1915+105 respectively). The strong, large scale
field in the upstream of the reconnection region that approaches the
reconnection layer at subrelativistic speed results in a very efficient
acceleration configuration.

In the next section, I describe the mechanism for particle
acceleration to ultra-high energies. This mechanism is applied to relativistic jets
in Sect. 3. Discussion is given in Sect.~4.  

\section{Particle acceleration in magnetic reconnection regions}

In this work, I do not assume any specific mechanism for magnetic
reconnection. I consider the rather generic geometry where magnetized fluid
with reversing polarity over a lengthscale $l_{\rm rec}$ is advected into the dissipation
layer with speed $\beta_{\rm rec}=v_{\rm rec}/c$. The plasma is heated/compressed
and accelerated by the released magnetic energy  
and leaves the region through a narrow layer of thickness $\delta\ll
l_{\rm rec}$ at the Alfv\'en speed of the upstream plasma $\beta_{\rm out}\sim \beta_{\rm
  A}$ (e.g. Lyubarsky 2005; see fig.~1). For high $\sigma$ flows
considered here $\beta_{\rm out}\sim 1$.

Particle acceleration at the reconnection layer can be complicated by
the exact, small-scale reconnection geometry. At the location where magnetic energy
is dissipated one may deal with current sheets (Parker 1957; Sweet 1958),
slow MHD shocks (Petschek 1964; Zenitani, Hesse \& Klimas 2009), 
MHD turbulence (Lazarian \& Visniak 1999; Matthaeus \& Lamkin 1986; 
Loureiro et al. 2009) and/or secondary
tearing instabilities (Drake et al. 2006; Loureiro, Schekochihin \& Cowley 2007; Samtaney et
al. 2009). Regardless of the details of the reconnection mechanism,
however, any successful reconnection model must explain the observed fast rate with which
magnetic field lines are advected into the reconnection region
(e.g. Lin et al. 2003). The so-called reconnection speed has to be a substantial fraction
(a tenth or so) of the Alfv\'en speed $\beta_{\rm A}$ of the upstream plasma.
For $\beta_{\rm A}\sim 1$ the reconnection speed is subrelativistic:
$\beta_{\rm rec}\equiv \epsilon \sim 0.1$.

I assume that a preacceleration mechanism is able to accelerate ions
to large enough energy so that their gyroradius $R_{\rm g}$ becomes larger than
the thickness of the layer $\delta$. Such preacceleration may take place in contracting magnetic
islands (Drake et al. 2006), current sheets (e.g. Kirk 2004) or slow MHD
shocks. After the initial acceleration (discussed in Sect. 3.3), the particle trajectories and
energy will be subject to at most mild change while they cross the dissipation layer. Still, as long 
as $R_{\rm g}\simless l_{\rm rec}$, the particles are confined in the
wider reconnection region and can be accelerated further.

\subsection{The acceleration cycle}

The acceleration of the particles in the reconnection region may be
viewed from two equivalent perspectives. In the rest fame of the
current layer, the upstream region contains an advective electric
field directed along the $x$ axis of strength 
$\mathcal{E}_{\rm x}=\beta_{\rm rec}B$  (the so-called reconnection
electric field).  A particle bounces back and forth around the
reconnection layer in a betatron-like orbit (also called Speiser
orbit) schematically shown in Fig.~1. It is continuously accelerated by the
electric field with its energy increasing linearly with distance $x$
traveled along the layer $E'\sim e\beta_{\rm rec}Bx$ (Speiser 1965). 
The acceleration may also be viewed as result of repeated magnetic 
reflections in the upstream flow. The fact that the upstream is 
converging towards the dissipation region, allows for a first order 
Fermi acceleration to operate (de Gouveia dal Pino \& Lazarian 2005). 
As a result, particles gain a fixed fractional energy per cycle
that can be estimated by the following considerations (that 
involve two Lorentz boosts).

Consider that a particle of Lorentz factor $\gamma_1\gg 1$ 
(measured in, e.g., the lower side of the upstream region; see Fig.~1)
enters at an angle $\theta$ (with respect to the normal to the
reconnection layer, i.e., the $y$ axis) into the opposite side of the upstream.
The relative speed of the two regions is 
\be
\beta_{\rm r}=\frac{2\beta_{\rm  rec}}{1+\beta_{\rm rec}^2}.    
\ee  
The Lorentz factor $\gamma_2$ of the same particle measured in the rest frame of the upper side is
given by the transformation $\gamma_2=\gamma_1 \gamma_{\rm r}(1+\beta_{\rm r}\cos\theta)$.
The particle performs a fraction of a gyration and leaves the
upper side at an angle $-\theta$.\footnote{In general the particle will
  return at an angle $|\theta'|$ different than $\theta$. The change in the angle is 
  small during a single cycle for subrelativistic $\beta_{\rm rec}$ discussed here and can
  be ignored. Furthermore, a small deflection of the particle may take place while
  it crosses the reconnection layer that introduces some degree
  stochasticity to the orbit but does not affect the main arguments
  presented here.} The particle completes the cycle by
returning to the lower region with Lorentz factor  (in the rest frame of that region)
$\gamma_3= \gamma_2 \gamma_{\rm r}(1+\beta_{\rm r}\cos\theta)$.
The amplification in energy during the $1\to 2\to 1$ cycle is 
\be
A(\theta)\equiv \frac{\gamma_3}{\gamma_1}=\gamma_{\rm r}^2(1+\beta_{\rm
  r}\cos\theta)^2.
\ee

%-----------------------------------------------------------------  
\begin{figure}
\resizebox{\hsize}{!}{\includegraphics[angle=90]{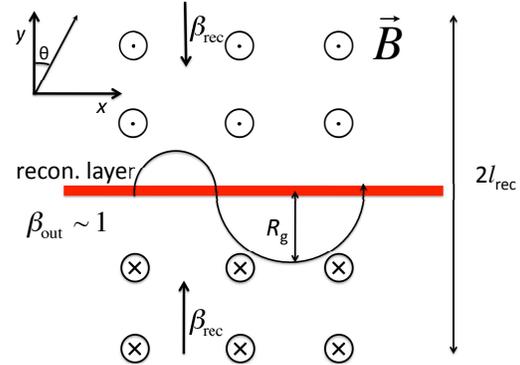}}
\caption[] {Sketch of the reconnection geometry. High-$\sigma$
plasma approaches the reconnection layer at subrelativistic speed
$\beta_{\rm rec}$ from both sides (upstream flow). Plasma leaves the region through a
thin layer with $\beta_{\rm out}\sim 1$ perpendicular to the plane of
the sketch (downstream flow). Energetic ions are Fermi accelerated by been 
repeatedly reflected by the upper and lower sides of the upstream. 
\label{fig1}}
\end{figure}
%----------------------------------------------------------------

In the limit of $\theta=0$ (particle crossing perpendicular to the
reconnection layer), the amplification $A$ is given as function of $\beta_{\rm rec}$:
\be
A(0)=\frac{1+\beta_{\rm r}}{1-\beta_{\rm r}}=\frac{(1+\beta_{\rm
    rec})^2}{(1-\beta_{\rm rec})^2},
\ee
where eq.~(1) is used in the last step.
If any scattering process (e.g. when crossing the reconnection layer) 
keeps the particle distribution quasi-isotropic, particles will cross 
the dissipation region with a angular 
probability distribution $P(\theta){\rm d}\theta=2\sin\theta\cos\theta
{\rm d}\theta$ with $0\le \theta\le \pi/2$.   
In this case, the average amplification per circle is
\be
<A>=\int_0^{\pi/2}A(\theta)P(\theta)
{\rm d}\theta=\gamma_{\rm r}^2[1+\frac{4}{3}\beta_{\rm
  r}+\frac{1}{2}\beta_{\rm r}^2].
\ee
Note that, in the limit $\beta_{\rm rec}\ll 1$, $<A>=1+8\beta_{\rm
rec}/3$ as found in de Gouveia dal Pino \& Lazarian (2005). 

In Fig.~2, I plot the $A(0)$ and $<A>$ as a function of $\beta_{\rm
rec}$. Note that $A(0)$ and $<A>$ have rather similar values making 
the results that follow rather insensitive to the exact angular
distribution of the particles. One can see that for reasonable values 
of $\beta_{\rm rec}\sim 0.2$, amplification of factor of $\sim 2$ is expected.

%-----------------------------------------------------------------  
\begin{figure}
\resizebox{\hsize}{!}{\includegraphics[angle=270]{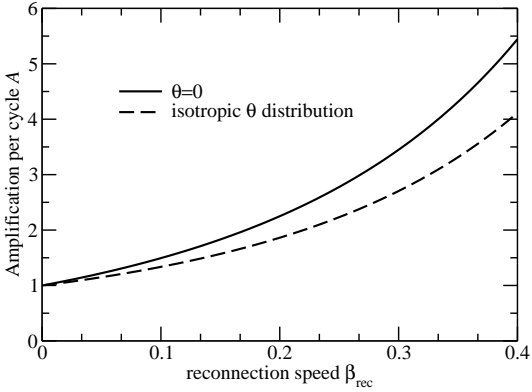}}
\caption[] {Energy amplification $A$ of particles for every Fermi acceleration
  cycle as function of the reconnection speed $\beta_{\rm rec}$. The solid
  curve corresponds to particles crossing the reconnection layer at
  an angle $\theta=0$ and the dashed curve to an isotropic particle distribution.       
\label{fig2}}
\end{figure}
%----------------------------------------------------------------

\subsection{Acceleration time scale}

One acceleration cycle for a particle of energy $\gamma$
approximately equals the  gyration time $t_g=2\pi \gamma mc^2/eBc$ at this
energy\footnote {The energy changes moderately during the cycle; we
consider the final energy of the particle during the cycle slightly
overestimating the gyration timescale and therefore the acceleration time.},
where $m$ is the mass of the particle and $B$ the magnetic field
strength of the upstream region. At the previous cycle the
particle has $A^{-1}$ less energy (and the cycle lasts proportionally shorter).
The total time (acceleration time) for a particle of Lorentz factor
$\gamma$ is the sum of the time spent in all the cycles until it reaches $\gamma$:
\be     
t_{\rm acc}=\frac{2\pi \gamma mc^2}{eBc}(1+A^{-1}+A^{-2}+...)=\frac{2\pi \gamma mc^2}{(1-1/A)eBc}.
\ee
For reasonable reconnection rates, the amplification is $A\sim 2$ which corresponds to
acceleration taking place at a timescale of the order of the gyration
period for the maximum energy  of the particles.

For a particle moving at a small angle with respect to the $x$
direction ($\theta\sim \pi/2$ case), the acceleration time scale can be
estimated by setting the energy of the particle $\gamma mc^2$ equal
to $e \beta_{\rm rec}Bx$, where $x$ is the distance
travelled along the layer. The acceleration time is, therefore,
$t_{\rm acc}\simeq x/c=\gamma mc^2/(e\beta_{\rm rec}Bc)$.
For $\beta_{\rm rec}\sim 0.1-0.2$, the previous expression results in
$t_{\rm acc}\sim t_{\rm g}$. This estimate is similar to that
of eq.~(5) showing that the $\theta$ distribution of the
particles has little effect on the timescale for particle
acceleration.  

\section{Constraints on the maximum particle energy}

A number of physical processes can limit the maximum energy that a particle
acquires. The acceleration ceases once the dynamical timescale of the system
becomes comparable to the acceleration time $t_{\rm acc}$ or when the
particle can no longer be confined within the reconnection region or when cooling 
of the particle is fast enough to inhibit further acceleration. Each of these processes is discussed in turn.  

In the following estimates, the magnetic field strength of the jet plays a
critical role. It is determined by the (Poynting) luminosity $L$, bulk Lorentz 
factor $\Gamma$ of the jet and the distance $R$ from the central engine:
\be B=\frac{L^{1/2}}{c^{1/2}R\Gamma}, \label{B}\ee where the magnetic
field $B$ strength is measured in the rest frame of the jet. For observed energy $E$ of a
UHECR, the energy of the particle in the rest frame of the jet is
$E'=E/\Gamma=\gamma mc^2$.
Throughout this work, we consider proton acceleration, i.e, $m=m_{\rm
  p}$ (the somewhat relaxed constraints
for acceleration of heavier nuclei are straightforward to derive). 

{\it Confinement Constraints:} for the particle to be confined within
the reconnection region, the gyroradius $R_{\rm g}$ cannot exceed the region's
thickness $l_{\rm rec}$.\footnote{This constraint may be somewhat
relaxed if the particles move predominately along the reconnection
layer ($\theta\sim \pi/2$) while been accelerated by the advective electric
field.} Setting $R_{\rm g}=l_{\rm rec}$ and using
eq.~(\ref{B}) I find that
\be
E_{\rm M}^{\rm conf}=\frac{el_{\rm rec}}{R}\sqrt{\frac{L}{c}}.\label{conf}
\ee
During their acceleration drift along the $x$ direction, the particles
may eventually leave the reconnection region. For a given length of the
reconnection layer $l_{\rm x}$ the maximum energy
that can be achieved is $E_{\rm M}=\Gamma e \beta_{\rm rec}Bl_{\rm
  x}=e\beta_{\rm rec}l_{\rm x}(L/c)^{1/2}/R$. For $l_{\rm x}\sim
l_{\rm rec}$, the last expression gives a maximal energy that
is a factor $\beta_{\rm rec}$ smaller than $E_{\rm
  M}^{\rm conf}$. On the other hand, in the model that is discussed 
in the next section, $l_{\rm x}$ is sufficiently larger than $l_{\rm
  rec}$ and the escape along the layer is no more constraining for 
the maximum energy than the estimate of eq.~(7).

Scattering processes can lead to particle motion perpendicular
to the plane of Fig.~1 (along the reconnecting magnetic field; 
i.e., along the $z$ axis). Escape of particles in the $z$ direction will depend on the
size of the layer $l_{\rm z}$ and the nature of the scattering
process. Particle escape in this direction is likely a stochastic 
process and may (in connection to the amplification per cycle $A$)
determine the distribution function of the accelerated 
particles but does not limit the maximum achieved energies.

{\it Cooling constraint:} the accelerated protons cool through
synchrotron emission and photo-pion production. Both of these 
processes can limit the energy of 
the UHECRs. In practise, however, it can be shown that
cooling is dominated by the synchrotron emission (Waxman 1995). 
The proton-synchrotron cooling timescale 
$t_{\rm syn}\simeq \frac{3m_{\rm p}^4c^7\Gamma}{e^4B^2E}$
can be set equal to the gyration time $2\pi E/ec\Gamma B$
to derive
\be
E_{\rm M}^{\rm syn}=\sqrt{\frac{3}{2\pi e^3B}}(m_{\rm p}c^2)^2\Gamma=
\sqrt{\frac{3R}{2\pi e^3}}(m_{\rm p}c^2)^2\Gamma^{3/2}\Big(\frac{L}{c}\Big)^{-1/4}, 
\label{cooling}
\ee   
where eq.~(6) is used in the last step of the derivation.

{\it Timescale constraint:} The time available for the acceleration
process to take place is limited by the dynamical (expansion)
timescale of the jet. Equating the time $R/\Gamma c$ that takes for
the jet to double its radius (in its rest frame) to the acceleration
time (taking it equal to the gyration time; see Sect. 2.2),
one finds
\be
E_{\rm M}^{\rm time}=\frac{e}{2\pi\Gamma}\sqrt{\frac{L}{c}}. \label{time}
\ee
In the next section, these constraints are applied to specific sources.

\subsection{GRB jets}

GRB jets have been proposed as sources of UHECRs by Milgrom \& Usov
(1995), Waxman (1995), and Vietri (1995) (see also Rieger \& Duffy
2005; Murase et al. 2006; 2008).
Using the observed long-duration GRB rate (e.g., Guetta, Piran \&
Waxman 2005) the energy release in $\gamma$-rays from bursts at the local
Universe can be estimated to be $\sim 10^{44}$erg/Mpc$^3$yr. This is
comparable to the rate of energy release required to
power the observed UHECRs with $E\simmore 10^{19}$ eV (Waxman 1995). 
GRBs are, thus, an energetically viable source {\it provided} that they can accelerate
UHECRs with similar efficiency to which they produce $\gamma$-rays.  

The (isotropic equivalent) $\gamma$-ray energy of long-duration 
GRBs is $E_{\gamma}\sim 10^{53}$ erg while their duration is $\sim
10$s. This corresponds to typical $\gamma$-ray luminosity
of $L_{\gamma}\sim 10^{52}$ erg/s.  During the active GRB phases, the 
luminosity of the jet is larger because of the likely moderate 
efficiency $\simless 50$\% in producing $\gamma$-rays, and the
quiescent intervals in between ejection events. Here, I conservatively
assume typical flow luminosity of $L\sim 10^{52}$ erg/s.
Furthermore, compactness arguments (e.g. Lithwick \& Sari 2001) bring
the bulk Lorentz factor of the flow into the hundreds: $\Gamma\sim 100-1000$. 

Normalising $L=10^{52}L_{52}$ erg/s, $\Gamma=10^{2.5}\Gamma_{2.5}$
and the dissipation radius (to be estimated below) to $R=10^{13}R_{13}$ cm,
the cooling and timescale constraints (\ref{cooling}) and
(\ref{time}) read $E_{\rm M}^{syn}=7\times 10^{19}
\Gamma_{2.5}^{3/2}R_{13}^{1/2}L_{52}^{-1/4}$ eV and
$E_{\rm M}^{time}=9\times 10^{19}L_{52}^{1/2}\Gamma_{2.5}^{-1}$ eV
respectively. The last expressions show that proton energies of
$E\sim 10^{20}$ eV are within reach if the magnetic reconnection takes 
place at large enough distance. 

The efficient dissipation of magnetic energy through reconnection
presumes that the jet contains regions with reversing magnetic 
fields. These field reversals may be imprinted in the flow 
from the jet launching region (i.e. one deals with 
a non-axisymmetric rotator at the central engine) or are developed further out in the flow
because of MHD instabilities. In the former case (i.e., that of an oblique rotator), the 
length scale of field reversals can be straightforwardly estimated to be (in the lab
frame) of order of the lightcylinder of the central engine 
$l^{\rm lab}_{\rm rec}\simeq c T_{\rm CE}$,
where $T_{\rm CE}$ is the rotation period of the central engine. 
In the rest frame of the flow, the field reversal takes place on a
scale $l_{\rm rec}=\Gamma l^{\rm lab}_{\rm rec}=\Gamma c T_{\rm CE}$
(see, e.g., Drenkhahn 2002). In this picture, reconnection takes place 
over a range of distances but is completed at a distance $R_{\rm rec}$ for
which the time scale of reconnection $l_{\rm rec}/\epsilon c$ equals 
the expansion timescale $R/\Gamma c$.
Assuming a millisecond period rotator as the central engine of the
GRB, the reconnection radius is (see Drenkhahn \& Spruit 2002 for
a more detailed derivation)
\be
R_{\rm rec}=\frac{\Gamma^2cT_{CE}}{\epsilon}=3\times 10^{13}
\frac{\Gamma_{2.5}^2T_{-3}} {\epsilon_{-1}}\qquad \rm cm. \label{Rrec}\ee 

Using the last expression for the dissipation radius {\it predicted}
by the reconnection model, one can write down the various constraints
(7), (8), and (9) as
\begin{eqnarray}
E_{\rm M}^{\rm conf}&=&6\times
10^{19}\frac{\epsilon_{-1}L_{52}^{1/2}}{\Gamma_{2.5}}\quad \rm eV, \nonumber\\
E_{\rm M}^{\rm syn}&=&1.2\times
10^{20}\frac{\Gamma_{2.5}^{5/2}T_{-3}^{1/2}}{L_{52}^{1/4}\epsilon_{-1}^{1/2}}\quad \rm eV, \\
E_{\rm M}^{\rm time}&=&9\times 10^{19}\frac{L_{52}^{1/2}}{\Gamma_{2.5}}\quad \rm eV \nonumber.
\end{eqnarray}
Interestingly, for $\epsilon\simeq 0.15$ the first and third constraints give
the same result. Proton energies up to $E\sim 10^{20}$ eV are plausible.

In this model, UHECRs are accelerated at a distance $R_{\rm rec}$
(see eq.~(10)) which is much shorter than the distance where the jet decelerates
interacting with the external medium. Adiabatic losses could,
therefore, reduce the energy with which the particles escape.
However, because of the efficient dissipation of Poynting flux
at $R_{\rm rec}$, the magnetic field strength drops with distance
steeper than $R^{-1}$ for $R\simmore R_{\rm rec}$. It is, therefore, reasonable 
to assume that at least the most energetic particles become
effectively decoupled from the jet and avoid adiabatic cooling.  
Note that in models for UHECR acceleration where the flow maintains a
fixed fraction of Poynting flux, the magnetic field strength scales as
$B\propto 1/R$ (see eq.~(6)). In those models, after the particle acceleration
is completed $R_{\rm g}/ct_{\rm dyn}\propto (RB)^{-1}=\rm{const.}$; 
even very energetic particles remain coupled to the 
flow potentially suffering severe adiabatic losses. 

\subsection{Jets in active galactic nuclei}

Powerful AGN jets can also potentially accelerate protons to UHECRs. Since we
are not aware of any such jet with, say, $L\sim 10^{48}$ erg/s 
within the Greisen, Zatsepin Kuzmin (GZK)  radius (Greisen 1966; Zatsepin 
\& Kuz'min 1966), one has to assume
that such sources are transient (see Waxman \& Loeb 2009 for
discussion on observational constraints on the existence of sufficient
number of AGN jets to power the observed UHECRs).
A powerful AGN jet of $L\sim 10^{48}$ erg/s and bulk $\Gamma\sim 3$
that contains field reversals on scales $l_{\rm rec}\sim cT_{\rm CE}$ with
$T_{\rm CE}\sim 10^5$ s (i.e. some 8 orders of magnitude longer than that
of the GRB central engine due to the larger mass of the compact
object) can also satisfy the basic constraints for 
proton acceleration to the highest observed energies.  For parameters
relevant for AGN jets, eq.~(11) gives:
\begin{eqnarray}
E_{\rm M}^{\rm conf}&=&6\times
10^{19}\frac{\epsilon_{-1}L_{48}^{1/2}}{\Gamma_{0.5}}\quad \rm eV, \nonumber\\
E_{\rm M}^{\rm syn}&=&1.2\times
10^{20}\frac{\Gamma_{0.5}^{5/2}T_{5}^{1/2}}{L_{48}^{1/4}\epsilon_{-1}^{1/2}}\quad \rm eV, \\
E_{\rm M}^{\rm time}&=&9\times
10^{19}\frac{L_{48}^{1/2}}{\Gamma_{0.5}}\quad \rm eV \nonumber.
\end{eqnarray}
It remains, however, to be shown that such powerful
jets are occasionally activated in nearby AGNs (see Farrar \& Gruzinov 2009
for a proposed mechanism for such AGN flares).

Unlike the GRB flow that is expected to consist of protons (and maybe
neutrons) because of the extreme temperatures and densities of the
launching region, AGN jets may contain heavier nuclei such as iron.
For iron composition of the UHECRs, the constraints for acceleration on the
luminosity of the source are significantly relaxed (see, e.g., 
Pe'er, Murase, M{\'e}sz{\'a}ros 2009; Honda 2009; Dermer \& Rozaque 2010). 
Magnetic reconnection in less powerful AGN jets of $L\sim
10^{44}-10^{46}$ erg/s can accelerate iron up to energies of $10^{20}$ eV. 

\subsection{Injection energy}

The particles that enter the acceleration cycle discussed here, have passed through
a preacceleration phase that makes their gyration radius $R_{\rm g}$
comparable to the thickness of the layer $\delta$. The required energy of the
preacceleration phase depends on the details of the reconnection
geometry but may be roughly estimated by the following considerations. Plasma
of density $\rho_{\rm in}$ enters the reconnection region with speed 
$\beta_{\rm rec}$ and leaves it with density $\rho_{\rm out}$
at the Alfv\'en speed $\beta_{\rm A}=\sqrt{\sigma/(1+\sigma)}$ 
through the layer: $\beta_{\rm rec}\rho_{\rm
  in}l_{\rm rec}=\gamma_{\rm A}\beta_{A}\rho_{\rm out}\delta$, where the length of the
layer is assumed $l_{\rm z}\sim l_{\rm rec}$. The compression
$\rho_{\rm out}/\rho_{\rm in}$ in the layer is model
dependent. Assuming, for example, a Petschek-type geometry $\rho_{\rm
  out}/\rho_{\rm in}\sim 2 \sigma^{1/2}$ (Lyubarsky 2005), the
thickness of the layer is $\delta\sim l_{\rm rec}\epsilon/2\sigma$.

The downstream magnetic field $B''$ (in the outflow frame) contributes
only a fraction to the total pressure (the rest coming from hot
particles). Pressure balance across the reconnection layer, therefore, constrains
$B''\simless B$ (for thermally dominated downstream $B''\ll B$). 
In the downstream frame, the preacceleration energy $E''_{\rm preacc}$ is 
given by setting $\delta=R_{\rm g}=E''_{\rm preacc}/eB''$ which yields
$E''_{\rm preacc}\sim eB''\delta\simless eB\delta \sim eB l_{\rm
rec}\epsilon/2\sigma$. For typical GRB parameters and $\sigma=100\sigma_2$,
$E''_{\rm preacc}\simless 10^{14}\epsilon_{-1}^2L_{52}^{1/2}\Gamma_{2.5}^{-2}\sigma_2^{-1}$ eV.
 This is to be compared with the thermal energy in the downstream:
 $E''_{\rm thermal} \sim \sigma^{1/2}m_pc^2\sim 10^{10}\sigma_2^{1/2}$
 eV.  Substantial preacceleration above the thermal energy may
 therefore be required for the
mechanism to operate. Tearing instabilities in the current layer 
are, however, expected to lead to fluctuations in the thickness of the 
layer $\delta$ and to regions where the preacceleration requirements are
substantially reduced leading to the bulk of the particle injection.    

\section{Discussion}

Magnetic reconnection results in directly {\it observed} energetically
substantial, non-thermal, high-energy particles during solar flares (Lin 
\& Hudson 1971; Lin et al. 2003) or in Earth's magnetotail (Terasawa \& Nishida 1976;
Baker \& Stone 1976; {\O}ieroset et al. 2002).
The puzzling particle acceleration measured by Voyager 1 and Voyager 2
{\it well} beyond the termination shock of the solar wind (Stone et al. 2005;
2008) may take place at reconnection regions in the shocked
solar wind (Lazarian \& Opher 2009; Drake et al. 2010). Beyond our solar system, 
magnetic reconnection may operate in a variety of relativistic
environments accelerating particles. Magnetic reconnection in
the pulsar winds (Coroniti 1990;  Lyubarsky \& Kirk 2001), the termination 
shock of pulsar winds (Lyubarsky 2003) or GRB flows 
(Thompson 2006) may result in particle acceleration and
emission from these sources. If MHD jets contain field reversals on
sufficiently small scale (of the order of the light cylinder), 
reconnection can efficiently power the GRB
(Spruit et al. 2001; Drenkhahn \& Spruit 2002; Giannios 2008) and 
AGN jet emission (Sikora et al. 2005; Giannios, Uzdensky \&
Begelman 2009, 2010; Nalewajko et al. 2010).

Here, I present a model for UHECR acceleration in magnetic
reconnection regions in strongly magnetized (AGN or GRB) jets. 
I argue that the converging magnetized plasma offers an ideal 
trap for particles to be accelerated to the highest observed energies. The mechanism
discussed here assumes that reconnection is fast $\beta_{\rm rec}\sim
(0.1-0.2) \beta_{\rm A}$ but does not depend on, poorly known, reconnection physics.
It is an efficient mechanism that accelerates particles on a timescale of
order of their gyration period. 
Protons can reach $\sim 10^{20}$ eV in GRB and luminous AGN jets.

\section*{Acknowledgments}
I thank L. Sironi, A. Spitkovsky, H. C. Spruit, D. A. Uzdensky for
important comments and suggestions. 
I acknowledge support from the Lyman Spitzer, Jr. Fellowship awarded by the
Department of Astrophysical Sciences at Princeton University.

\end{document}